# Research trends in structural software complexity

Tom Mens

***Abstract***. *There are many dimensions of software complexity. In this article, we explore how structural complexity is measured and used to study and control evolving software systems. We also present the current research challenges and emerging trends in this domain that has remained active for nearly four decades, and continues to evolve.*

## 1 Introduction

Software systems are among the most intellectually complex artefacts ever created by humans [Brooks1986]. But what is it that makes software systems complex? Despite several decades of research on the topic, there is no consensus on how to define software complexity. The IEEE standard 610.12-1990 defines software complexity as "the degree to which a system or component has a design or implementation that is difficult to understand and verify". Basili defines complexity in a more general way as "*a measure of the resources expended by a system while interacting with a piece of software to perform a given task*" [Basili1990]. Basili's definition reveals that there are many different dimensions of complexity.

If the system that interacts with the software is a *computer*, then complexity is defined by the execution time and memory resources required to perform the computation. We call it *algorithmic complexity* if we are concerned by the time and storage required to execute a particular algorithm. We call it *computational complexity*, if we are interested in the theoretical complexity or difficulty (in time or space) of the computational problem to be solved, regardless of the algorithm used to solve it. Algorithmic and computational complexity theory are very important and well-developed, with lots of theoretical models of computation (e.g. Turing machines, recursive functions) and complexity classes (e.g. polynomial time complexity, logarithmic space complexity, and so on). One could argue that computational complexity corresponds to what Fred Brooks calls the *essential complexity*, which is intrinsic to the characteristics of the problem to be solved and therefore cannot be reduced. The algorithmic complexity then relates to Brooks' notion of *accidental complexity*, which is introduced by the way in which the problem is implemented. It depends on the chosen algorithm, but also on the software engineering tools used to implement the problem (such as programming language, compilers and other technical characteristics). Brooks argues that the essential complexity cannot be reduced, and removing accidental complexity (e.g. by using higher-level programming languages) can never lead to an order of magnitude productivity improvement.

If the interacting system is a *user*, then complexity is defined by the difficulty of using the software, which can be expressed as the number of different ways the user can or needs to interact with the software. This *complexity of use* refers to the software quality characteristic of *usability*. It is also related to the *functional complexity*, i.e., the amount of functionality the software provides to the user [TranCao2001]. Albrecht's *function points* metric [Albrecht1979] and related functional size measurements have been widely used to measure this kind of complexity.

If the interacting system is a *software developer*, then complexity is assessed by the difficulty of performing tasks such as coding, debugging, testing or modifying the software. We refer to this type of complexity as *structural software complexity* [Darcy2005]. It reflects how the software is structured and organized in different interacting parts (both at fine-grained and coarse-grained level), because this structure has a direct effect on other quality characteristics such as understandability and maintainability. The correlations between software complexity metrics and maintenance aspects have been widely studied. Strong relationships have been found between complexity and increased maintenance effort, number of bugs and amount of changes [Kemerer1995]. Large chunks of "spaghetti code" require more effort to read and maintain than well-modularized software composed of weakly coupled modules. Structural complexity is different from computational complexity because it focuses on the solution (the software itself) as opposed to the computational problem to be solved. It is also different from the algorithmic complexity, which focuses on the machine resources (time and memory consumption) required to solve the problem.

Finally, if the interacting system is the *organization or team producing the software*, then the complexity is defined by the way in which team members interact, collaborate and communicate to develop and maintain the software product over time. We will refer to this as the *organizational complexity* [Bird2009b]. Such organizational complexity can be an important cause of structural software complexity: managing large, possibly distributed, development teams is challenging, with specific problems such as the communication overhead among members of the development team, or an overly complex organizational

---

**Some dimensions of Software Complexity**

**Theoretical complexity:** The complexity of the problem to solve, in terms of time and space.
- Computational: inherent to the problem
- Algorithmic: of a particular solution algorithm

**Complexity of use:** How hard is it to use the software?
- Functional: Number and difficulty of functionalities
- Usability: Difficulty of interacting with the software

**Structural complexity:** How hard is it to for a developer understand and maintain the software?
- Module-level
- System-level

**Organisational complexity:** How hard is it for the development team to coordinate and collaborate on the development and maintenance of the software?

structure. As a consequence, organizational complexity should be considered together with structural complexity measures to estimate software quality and to predict failure-proneness and other software problems.

Our main focus of will therefore be on structural and organizational complexity, since these appear to be the most relevant from a project management point of view. According to Darcy, structural complexity involves intellectual resources (programmer understanding and effort) that do not increase over the years as opposed to computing resources that follow Moore's law and its successors [Darcy2005]. Along the same line, Midha showed that higher structural complexity coincides with more bugs, increased maintenance effort an increasing difficulty of incorporating new developers [Midha2008]. Therefore, dedicating efforts to controlling and reducing complexity is crucial for successful projects.

The remainder of the article will be structured as follows. We start by presenting a short historic overview. Then we focus on the different ways to measure structural complexity at the level of software modules, as well at the level of the module dependency graph, and we discuss how to aggregate module-level measures to obtain a system-level complexity measure. Next, we explore the evolution of structural complexity as the software system evolves over time, and report on some empirical studies that have gained insight in this evolution phenomenon. Finally, we address organizational complexity and how it relates to the structural software complexity and its evolution.

## 2 A short history

Research on software complexity has been around since the 1970's, with the introduction of several new software metrics that measure some aspect of software complexity [McCabe1976, Halstead1977, Albrecht1979]. Since the 1980's research has focused on evaluating and comparing these metrics (e.g., [Henry1981, Henry1984, Kafura1985, Kafura1987]). Since the 1990's, with the emergence of the object-oriented programming paradigm, OO metrics emerged, of which the C&K suite is undoubtedly the most important one that is still widely used today [Chidamber1994]. Since the 2000's, there was a renewed focus and interest in software evolution research inspired by Manny Lehman's laws of software evolution and the agile software development methodologies, and facilitated by the emergence and proliferation of freely available open source software repositories from which large evolution histories of software projects could be extracted [MacCormack2006, Midha2008, Sangwan2008, Bird2009a, Bird2009b, Darcy2010, Terceiro2010]. This research still goes on today, but now also considers the importance of organizational and social aspects [Nagappan2009], combined with insights and techniques from network science and complex systems research [Valverde2002, Myers2003, Potanin2005, Louridas2008] and economic aggregation measures [Vasa2009, Serebrenik2010].

## 3 Measuring structural software complexity

[Brooks2003] considers the lack of a good metric of structural complexity as one of three great challenges in computer science: "We have no theory that gives us a metric for the information embodied in structure. […] We cannot say it is $x$ times more complex where $x$ is some number.'' The first step in taming structural complexity is measuring it ("You get what you measure"). As soon as a measure for a goal is defined, the metric will move towards its desired value, because the team will start working to reach this goal. By measuring complexity one may get to understand its causes, to detect which parts of the system are a potential source of current or future maintenance problems and therefore, to determine what should be changed and how [Kafura1987].

To measure the structural complexity of software, one should distinguish between approaches aiming to assess the complexity of the software system as a whole, and those aiming to measure the complexity of individual software modules[1]. The former tend to rely on the graph of module dependencies and notions such as coupling and cohesion, while the latter rely on a detailed analysis of the control-flow or data-flow structure of a module.

### 3.1 Measuring complexity at system level

Many useful structural complexity measures exist. A low-level but easy to compute measure is the system's *size*. It is typically measured by counting the number of lines of code (LOC), but one can also measure the number of higher-level programming constructs. In practice, size is a good first indicator of a software system's complexity: today it is not unusual to find systems containing several millions of lines of code. However, measuring size alone is not sufficient: when comparing two systems of similar size, their complexity may be perceived as significantly different.

*Coupling* and *cohesion* are two different yet interrelated ways to measure the structural complexity of a software system. By relying on a theoretical task complexity model, Darcy *et al.* claim that coupling and cohesion are essential aspects of structural complexity [Darcy2005]. In addition, they can be defined for any programming paradigm (e.g., procedural or object-oriented programming) and programming language. Coupling and cohesion are defined at the level of a program module. *Coupling* of a module relies on the essential characteristic of "relatedness", as it is computed as the number of other modules to which the module refers. *Cohesion* of a module relies on the notion of "togetherness", and is expressed in terms of how the subparts of the module interact with each other. Specifically for the object-oriented paradigm, Chidamber and Kemerer [Chidamber1994] have proposed the LCOM metric (Lack of Cohesion of Methods in a class) as a way to measure the absence of cohesion, and two metrics CBO (Coupling

---

[1] In the remainder of this paper we will use the term "module" to refer generically to any kind of program unit, such as components, classes, methods, packages, procedures, functions, and so on.

Between Object classes) and RFC (Response for a Class) as a way to measure coupling. Various researchers later improved the LCOM metrics. [Purao2003] lists dozens of cohesion and coupling measures for the object-oriented paradigm, many of which are based on the original C&K metric suite. [Sangwan2008] showed that excessively complex systems tend to have high coupling and low cohesion. A module with a low coupling is largely independent of the other modules of the system. Darcy *et al*. [Darcy2005] carried out a controlled experiment with software professionals having several years of experience. They studied the influence of the level of coupling and cohesion on the effort to carry out perfective maintenance tasks. They concluded that coupling and cohesion have to be considered jointly as different but interdependent notions of complexity. They also found statistical evidence that for more highly coupled programs, higher levels of cohesion reduce maintenance effort.

All of the above approaches compute coupling by analyzing the syntactic relationships between modules in the program code. In addition to this, it is possible to find *implicit coupling* between modules that have no explicit structural relation between them. For example, Gall *et al*. extract "*logical coupling*" between modules that tend to change together, by analyzing of the evolution history of a system and its modules [Gall1998]. By taking into account such implicit dependencies, it may be possible to come to more comprehensive complexity measures.

The notion of coupling is essentially based on the dependencies between program modules. One can distinguish between the number of incoming dependencies (known *as fan-in* or *afferent coupling*) and outgoing dependencies (known as *fan-out* of *efferent coupling*). Henry and Kafura [Henry1984] defined the complexity of a module in terms as the square of the product of its number of incoming and outgoing dependencies. Kitchenham referred to this as *information flow complexity*, and studied its effectiveness in identifying software quality characteristics such as change-proneness, fault-proneness and subjective complexity as perceived by the developer [Kitchenham1990]. They concluded that simple code metrics such as LOC and number of branches performed better. Nevertheless, information flow metrics may still be useful since they are available earlier in the life-cycle. When focusing on fan-in and fan-out in particular, it turns out that they measure quite different things. Kitchenham found fan-out to be related to the aforementioned quality characteristics, while fan-in was not. [Troy1982] found similar results. In fact, modules with a large fan-in tend to be relatively small, hence, less complex, facilitating their reuse in many different contexts.

To get an idea of the global structural complexity of a software system, one can combine the incoming and outgoing dependencies of all modules, in order to obtain a directed *module dependency graph* representing the system structure. This dependency graph should preferably adhere to the *acyclic dependency principle*. Dependency cycles should be avoided, as they imply that all modules contained in the cycle depend on one another, making these modules harder to understand and modify.

---

**Some definitions of structural complexity metrics**

CBO(c) = coupling between object classes =
> the number of efferent couplings (i.e., fan-out) for class c. This coupling can occur through method calls, field accesses, inheritance, arguments, return types, and exceptions

RFC(c) = response for class c =
> the number of methods in the response set of a class, i.e., those methods that can potentially be executed in response to a message received by an object of that class.

LCOM(c) = lack of cohesion of methods in class c =
> the difference between the number of method pairs in a class that do not share some of the class's fields and the number of method pairs that do share a field access.

Information flow complexity =
> (fan-in * fan-out)$^2$

McCabe's cyclomatic complexity =
> the number of decision points in a control flow graph or, stated differently, the number of different execution paths that can be followed.

Halstead's effort E = D * V where
> program difficulty D = $(n_1 \times N_2) / 2n_2$
> program volume V = $(N_1+N_2) \times \log_2(n_1+n_2)$
> with $n_1$ the number of unique operators, $n_2$ the number of unique operands, $N_1$ the total number of operators and $N_2$ the total number of operands.

dep-degree (data-flow metric) =
> number of edges in the graph whose nodes represent program statements and whose edges relate variable definitions to variable uses.

---

### 3.2 *Measuring complexity at module level*

If the goal of a software developer is to get an idea of the effort required to understand or modify an individual software module, it often does not suffice to know how this module is related to, or coupled to, the other modules. Therefore, more fine-grained software metrics have been proposed to understand the internal structural complexity of software modules, primarily based on their control-flow structure or data-flow structure.

One of the first metrics of this kind to be proposed was McCabe's *cyclomatic complexity* [McCabe1976]. It measures the control-flow complexity at the level of functions, methods or procedures. The higher the cyclomatic complexity, the more difficult it will be for the developer to understand and modify the code. Gill and Kemerer [Gill1991] suggested that normalizing the cyclomatic complexity by LOC ("complexity density") is a more useful explanator of software maintenance productivity. Kafura and Canning [Kafura1985] suggested that information flow complexity is a better predictor than McCabe for the error-proneness of modules.

In a similar way, Shao and Wang defined a complexity metric belonging to the field of cognitive informatics [Shao2003]. They called it cognitive functional size (CFS). It claims to measure the difficulty of program comprehension, and like cyclomatic complexity it is based on the control-flow graph of a program. Based on empirical evidence it assigns different weights to different control structures of a program (e.g., iteration, sequencing, branching, function calls) to reflect their relative complexity.

Whereas the former two complexity metrics rely on the control-flow graph of a program, some metrics rely on the data-flow graph (involving variable definitions and variable uses) to measure structural complexity. An example is the dep-degree metrics proposed in [Beyer2010].

As a final example of popular module-level complexity metrics, Halstead defined the *effort* required to implement a program as the product of the program *volume* and the program *difficulty* [Halstead1977].

Many of these module-level metrics are correlated. For example, [Henry1981] showed that the Halstead metrics and McCabe's cyclomatic complexity are strongly correlated to each other, and to the lines of code (LOC) metric.

In order to obtain a system-level view on the structural complexity, module-level metrics need to be *aggregated* into a single value reflecting the structural complexity of the system as a whole. It is common practice to use central tendency measures (such as mean, standard deviation, weighted mean, median) as a way to aggregate the structural complexity over all modules. However, using these measures can be misleading: they may have an undesirable smoothing effect, may not be robust to outliers, and are known to be unreliable in presence of highly-skewed distributions. Such heavy-tailed distributions are common to most of the metrics that we have reported upon earlier, and economic inequality indices adopted from the field econometrics have been proposed and used to aggregate software metrics [Vasa2009, Serebrenik2010]. For example, the Gini, Theil, Atkinson, Hoover, Kolm index and related indices measure the inequality among values of a frequency distribution. The use of inequality indices as aggregation functions for module-level complexity metrics also presents some difficulties because they cannot discriminate between all values being equally low and all values being equally high. Because of this, a system with all modules being equally complex will have the same inequality index as a system with all modules being equally simple. In practice, however, this is unlikely to occur.

### 3.3 Discussion

In order to assess the structural complexity of software, one first needs to decide whether only a system-level view on complexity is needed, or whether the complexity at module level needs to be known as well. For example, if the goal is to decide whether a major restructuring of the software architecture is required, it probably suffices to analyze the module dependency graph for the presence of structural anomalies such as dependency cycles and non-respect of the principle of low coupling and high cohesion. If the goal is to assess the understandability and maintainability of individual modules with the aim to make local changes or perform local refactorings to reduce complexity, module-level metrics based on control-flow or data-flow are probably more useful. Based on these local metrics, one can still obtain system-level complexity values by aggregating results, but one should take care with aggregation functions such as mean and median because many metric distributions tend to be highly skewed. Economic inequality indices may provide a better alternative in these cases.

In any case, blindly applying a suite of complexity metrics is not recommended. There are too many such metrics (see, e.g. [Kafura1987, Chidamber1994]) that are often not orthogonal or measure entirely different things, and there are many challenges and limitations that one should address before using them in practice. Since many metrics are correlated, it should suffice to take a subset of them. However, it is still an open question as to what would be the most appropriate subset based on the characteristics of the system under study. Also, in some cases, a combination of different metrics may reveal statistical evidence that would not be observable by looking at the individual metrics in isolation. For example, Darcy *et al.* [Darcy2005] experimentally verified that neither coupling nor cohesion alone are sufficient to capture structural complexity at module level. Both notions are interdependent and need to be considered jointly when designing, understanding or maintaining software in order to effectively control its structural complexity. From this, it can be concluded that, when designing software, coupling and cohesion should be jointly considered to achieve the most desirable software structure.

## 4 Emerging trends

### 4.1 Laws of software evolution

Large software systems are very particular, and distinct from most other human-engineered system. They have been designed and organized with two, sometimes conflicting, goals in mind: to offer a given functionality to the user, and to be highly evolvable in order to accommodate for future changes. Developers continuously change the software in order to fix bugs, add new functionality, accommodate change requests, and to adapt the software to a changed environment. These changes are not seldom caused by a certain degree of uncertainty: a software system is originally conceived on the basis of incomplete or underspecified requirements, and needs to be adapted as the requirements change over time. These changes are unpredictable as they are triggered by events that are external to the system: users that find new ways to use the system for doing things it was not originally intended for; new technology and legislation to which the system needs to adapt; and so on. All of this contributes to the software's aging and erosion over time.

This led Manny Lehman to postulate his famous laws of software evolution. One of them is the law of *increasing complexity*: "as a program is evolved its complexity increases unless work is done to maintain or reduce it." This law is accompanied by other empirically validated laws of software evolution, such as *increasing growth* and *declining quality*. The relation between these laws can be intuitively explained as follows: in order to maintain customer satisfaction, more functionality is added to the system, leading to a growth of the system. With this growth comes an increase in interaction and dependencies between the system elements, leading to an increase in complexity. If this growth in complexity is not constrained, the effort needed to maintain the system becomes increasingly more important. As a result, the system will suffer from a declined quality: reduced reliability (more bugs), lower user satisfaction, and reduced flexibility of the system to adapt to future changes.

Numerous empirical studies have been carried out to understand how the complexity of software evolves over time, in order to find evidence or counter-evidence for the laws of software evolution. Some of these studies were carried out on proprietary software systems, while others analyzed open source software systems. Manduchi and Taliercio studied how software complexity evolves across releases for a proprietary Java software application [Manduchi2002]. They found that its complexity increased over time and no restructuring took place. On the other hand, [MacCormack2006] found that a redesign of the open source Mozilla web browser resulted in a reduced complexity as measured by coupling, and further evolution of the system maintained a low level of coupling despite a growth in size of the system. It hence seems that Lehman's law of increasing complexity is not always applicable. Darcy *et al.* explored this issue in depth by studying the evolution of size and structural complexity (measured as the product of coupling and lack of cohesion) for 108 relatively small open source projects taken from SourceForge [Darcy2010]. They classified these projects in three different clusters based on the observed evolutionary patterns. A cluster of 53 projects corresponded to an unchanged structural complexity over time, and coincided with a lack of growth in size for 43 of these projects. A second cluster of 38 projects had an increasing structural complexity. Finally, a third cluster of 17 projects revealed a decreasing structural complexity, while 11 of these projects experienced a growth in size. This seems to counter Lehman's law of increasing complexity, and illustrates that the evolution of size can behave differently from the evolution in structural complexity. Hence size and complexity metrics cannot be used interchangeably, as they reveal different evolutionary patterns for some projects. The authors also found that, over time, all projects tend to converge to a similar value of structural complexity. This may imply that there is some intrinsic level of complexity that projects achieve over time.

## 4.2   Inequality indices

Vasa *et al.* analyzed the structural complexity over time across a wide range of Java and C# software systems [Vasa2009]. Because many software metrics tend to have highly skewed non-Gaussian distributions, they used the Gini inequality index instead of central tendency measures such as mean or median. They observed that all project releases displayed remarkably high Gini indices, reflecting a highly unequal distribution of metrics values, but these values remained remarkably consistent over time. This seems to suggest that there may be a certain equilibrium in the degree of inequality that corresponds to a sustainable level of software evolution.

Serebrenik carried out a similar study, but using another economic inequality index, the Theil index [Serebrenik2010]. Interestingly, the Theil index is directly related to the notion of redundancy in information theory. The amount of redundancy can be seen as a measure of data complexity: the more noisy (or random) the data is, the less redundant (and more complex) it will be. This is the principle behind the concept of Kolmogorov complexity. It measures the information content of data and is complementary, yet deeply related, to the notion of Shannon entropy. A practical way of employing measures based on the Kolmogorov complexity is by using data compression algorithms and computing the normalized compression distance. Arbuckle has applied this idea to find more evidence of Lehman's law of increasing complexity, as the compression distance allows to abstract away of specific details of the software system (such as programming language used), focusing only on the information content of the software [Arbuckle2011].

## 4.3   Refactoring and restructuring

The effect of software refactoring and restructuring techniques on the evolution of structural software complexity is another aspect that requires further study. To analyze the effect of refactorings on the maintainability of software, [Kataoka2002] used a combination of coupling metrics and demonstrated how certain refactoring processes improved maintainability by reducing these complexity metrics.

[Sangwan2008] tracked the evolution of the structural complexity of three open source Java software projects. Although these projects had different evolution patterns of their structural complexity over time, they had something in common: high complexity shifted from lower levels (method and class level) to higher levels (package level) when carrying out refactorings to reduce the local complexity. Conversely, restructurings aiming to reduce a high global complexity shifted the complexity to the local level.

With the objective of reducing structural complexity, many different approaches have been proposed to suggest the most appropriate refactorings and restructurings. For example, combined metrics can be used as fitness functions for search-based procedures that find the best refactorings for a particular desired complexity optimization [Seng2006]. As another example, relational concept analysis has been used to compute the best way to remove a "god class" design smell by improving coupling and cohesion [Moha2008].

Nevertheless, while it seems that local refactorings result in local improvements, higher-level refactorings, to improve software structure at a higher level, should be planned. In order to widely improve software structure and therefore, to reduce system-level software complexity, one should perform strategic refactoring: high-level refactorings with a certain objective in mind [Neill2006].

## 4.4   Complex network analysis

The software development and maintenance process is built up from small local design steps carried out, often in parallel, by individual developers based on simple design principles such as decoupling, encapsulation, modularization, design patterns and refactoring. Nevertheless, these small steps lead to a *module dependency graph* with interesting complex network properties that *emerge* as a side effect of the development process without any particular software design principle that explicitly dictates this structure. Examples of such emerging properties are a *scale-free* topology and a *small-world* structure. Scale-freeness is observed if the proportion of nodes P(k) having k dependencies decays according to a *power law*[2]. The module dependency graph has a *small-world* structure if the average path length between any two nodes is very small and there is a large amount of clusters.

Many researchers have studied these emerging properties in module dependency graphs. Myers observed these properties in class dependency graphs of object-oriented systems [Myers2003]. Distinguishing between the incoming and outgoing module dependencies, he observed scale-free, heavy-tailed power-law distributions, with a higher exponent for out-degree than for in-degree, and a strong separation between nodes with large in-degree and nodes with large out-degree. Myers also studied how these complex network structures evolved over time. By analyzing the evolution of class dependency graphs over successive releases of a software system, he found that classes with a large out-degree tend to evolve more rapidly than classes with a large in-degree. A possible explanation could be that classes with high in-degree are highly reused by other classes and thus more constrained to remain stationary, while classes with high out-degree are typically bigger and more complex and thus more prone to changes.

Potanin *et al.* examined the structure of object dependency graphs, representing the execution (as opposed to the structure) of both procedural and object-oriented programs and they observed a similar kind of power-law distribution [Potanin2005]. Louridas *et al.* confirmed these results over a wider range of software systems, ranging over software systems, libraries, software distributions and compiled code, and including both object-oriented and procedural programs [Louridas2008]. Very similar power laws were observed for all of these systems, with a higher exponent for out-degree than for in-degree. Concas *et al.* not only observed the distribution of in-degree and out-degree in class dependency graphs but, they also investigated the distribution of various counting metrics [Concas2007]. In nearly all cases, they found evidence of a heavy-tail distribution, corresponding either to a log-normal distribution or a power-law distribution. In practice, it is often impossible to distinguish between these two types of distribution.

A possible explanation for these emerging patterns is that software design can be considered as an optimization process involving different conflicting objectives (e.g., reusability and maintainability versus performance and usability) and conflicting stakeholders (e.g., developers, managers and end-users). As such, the complex network topologies may well provide the best trade-off between high specificity (in order to implement the desired software functionality) and high evolvability (through code refactoring). According to Louridas, the observed structure in the module dependency graphs lies somewhere midway between optimal and worst-case structures. Software designers seem to perform following common sense based on their experience, rather than with a conscious effort of building scale-free structures.

In an attempt to formalize the above intuition, Myers and Concas *et al.* proposed models that may explain the process leading to complex network topologies. Myers simulated an oversimplified refactoring-based model of software evolution, and showed that the resulting complex network has properties that are very similar to those observed in real software systems. Concas *et al.* suggested to stochastically model random additions with the so-called Yule process, and showed that it lead to power law distributions. Many other models may lead to the same kind of emerging properties, so it still remains an open challenge to come up with models that most naturally resemble the software development process.

### 4.5 *Organizational complexity and socio-technical networks*

*Organizational complexity* refers to how a software-producing organization or development team is structured and the ease with which members of the software development team collaborate and communicate effectively and efficiently. In 1968, Melvin Conway observed that the structural system complexity mirrors the organizational complexity: "… organizations which design systems … are constrained to produce designs which are copies of the communication structures of these organizations". Conway's law implies that, in order to avoid software systems becoming overly complex, the organizational structure of the community developing the software may need to be changed and improved. This insight has recently lead to a new field of research known as *socio-technical network analysis*. Socio-technical networks combine module dependency graphs with contribution networks that represent which developer contributed to which software module. Socio-technical networks have been used, for example, to find evidence of Conway's law [Amrit2004]. Other researchers have contributed by using such networks to study or improve the communication between developers or to study which types of networks are most optimal (in terms of productivity, time needed for implementing a change request, reduction of software defects and so on).

Research in socio-technical network analysis has borrowed a series of well-known measures from social network analysis (e.g., betweenness centrality, closeness centrality, clustering coefficient). A challenge is to use these measures to improve the way in which to assess and control software complexity. Bird *et al.* [Bird2009b] applied these metrics to module dependency graphs, contribution networks and socio-technical networks combining the two former ones. Based on a validation in an industrial setting (on Windows Vista) and an open source setting (Eclipse) they found a significantly higher correlation with post-release defects when using socio-technical networks than when using dependency graphs alone. The correlation was also significantly higher than when using traditional code complexity metrics such as cyclomatic complexity. This indicates that it is important to combine information from social networks with structural information about the software to increase the predictive power of defect prediction models. It is quite possible that similar improvements may be found for other purposes as well.

---

[2] A distribution is scale-free if it is the same whatever the scale of observation. Power-law distributions satisfy this requirement.

MacCormack *et al*. studied whether there is a difference in how commercial software companies and open source communities develop software, and how this difference manifests itself in the software structure [MacCormack2006]. Strong supporting evidence was found for Conway's law. For loosely-coupled and highly distributed open source development teams, the software products were significantly more modular than for tightly coupled organizational structures of commercial software companies.

Terceiro *et al*. similarly studied how the level of project participation of open source contributors affects structural complexity, as measured in terms of coupling and cohesion [Terceiro2010]. By empirically analysing the open source Gnome projects, they found evidence that so-called core developers introduce less structural complexity than peripheral developers, and that core developers remove more structural complexity than peripheral developers.

Using Windows Vista as a case study, Bird *et al*. [Bird2009a] examined the effect of distributed development, within the same company and project, on software quality in terms of post-release failures. The authors found no significant difference in the failure rate between those components that were developed in a distributed fashion and those that were developed by collocated teams. Other component characteristics, such as code complexity, also differed very little between distributed and collocated components. An explanation can be that organizational issues are a much more important factor for software quality than geographical issues. To ensure good software quality, organization within a project has to be compact and follow some recommendations, like consistent usage of development tools, good relations between the different development sites, efforts to remove cultural barriers, synchronous communication, code ownership, common schedules and organizational integration.

## 5 Conclusion

We started this article by presenting the different dimensions of software complexity, and zoomed in on the notion of structural software complexity and how this evolves over time. We presented the state-of-the-art in research, distinguishing between module-level metrics based on control-flow and data-flow, and system-level metrics based on module dependency graphs. These traditional metrics have seen widespread use and have been relatively successful in models for predicting software defects and estimating software effort.

We discussed a number of current and future research avenues in structural software complexity. We suggested the use of economic inequality indices (as opposed to central tendency measures such as mean and median) to aggregate module-level complexity metrics to obtain a global system-level value. This was motivated by the fact that the metrics are typically not normally distributed but follow some kind of power-law distribution. This type of heavy-tailed distribution is often a sign of a scale-free small-world network topology, implying that it can be useful to use complex networks analysis for modeling and studying the evolution of software systems.

We also stressed the importance of not considering the structure of a software system in isolation, but to consider it as a socio-technical network. This implies that, in order to assess the software quality and complexity, one needs to take into account the organization, communication and interaction of the development team responsible for maintaining the software. For this, measures and techniques borrowed from project management and social network analysis can be exploited.

Another point of attention is the use of anti-regressive techniques to counter the effect of increasing complexity as the software system evolves and grows in size. Techniques such as refactoring and restructuring aid in this process, but more research is needed to understand how these techniques can be exploited in the most optimal way.

All of these techniques together will allow researchers to gain a better understanding of, and control over, how and why software evolves and becomes more complex over time. Such understanding will allow developers and managers to better manage software complexity, e.g., by improving the software structure itself, by changing the development process, or by changing the organization of the software development team.

# APPENDIX: ADDITIONAL REFERENCES